\documentclass[oribibl]{llncs} 
\usepackage[square,numbers]{natbib}

\usepackage[utf8]{inputenc} 
\usepackage[T1]{fontenc}    
\usepackage{lmodern}
\usepackage{hyperref}       
\usepackage{url}            
\usepackage{booktabs}       
\usepackage{amsfonts}       
\usepackage{amsmath}
\usepackage{nicefrac}       
\usepackage{microtype}      

\usepackage{graphicx}
\usepackage{color}
\usepackage{mathtools}

\usepackage{algorithm}
\usepackage{algorithmic}

\usepackage{caption}

\DeclarePairedDelimiter\floor{\lfloor}{\rfloor}

\newcommand{\trans}{\top} 

\newcommand{\DNorm}{\mathcal{N}}

\newcommand{\evid}{Z}

\newcommand{\iw}{W}  
\newcommand{\Riw}{\widetilde{\iw}}  

\newcommand{\psize}{N} 

\newcommand{\targd}{\mathbf{\pi}} 
\newcommand{\targdim}{D} 
\newcommand{\propd}{q} 

\newcommand{\SaS}{\mathbf{X}} 
\newcommand{\Wset}{\mathbf{W}} 
\newcommand{\RWset}{\widetilde{\Wset}} 
\newcommand{\smp}{{X}} 
\newcommand{\Rsmp}{\widetilde{\smp}} 
\newcommand{\RSaS}{\widetilde{\SaS}} 

\newcommand{\param}{\theta}

\newcommand{\PDK}{\mathbf{k}} 

\mainmatter              
\title{Kernel Sequential Monte Carlo}
\titlerunning{Kernel Sequential Monte Carlo}
\author{Ingmar Schuster$^\star$\inst{1} \and 
Heiko Strathmann$^\star$\inst{2} \and
Brooks Paige\inst{3} \and
Dino Sejdinovic\inst{4}}

\institute{FU Berlin, \email{ingmar.schuster@fu-berlin.de}
\and
Gatsby Unit, University College London, \email{heiko.strathmann@gmail.com}
\and
Alan Turing Institute and University of Cambridge, \email{bpaige@turing.ac.uk}
\and
University of Oxford and Alan Turing Institute, \email{dino.sejdinovic@stats.ox.ac.uk}}

\begin{document}

%

%

\maketitle

\renewcommand*{\thefootnote}{\fnsymbol{footnote}}
\footnotetext[1]{Equal contribution}
\renewcommand*{\thefootnote}{\arabic{footnote}}
\setcounter{footnote}{0}

%
%
%

\begin{abstract}
We propose kernel sequential Monte Carlo (KSMC), a framework for sampling from static target densities.
KSMC is a family of sequential Monte Carlo algorithms that are based on building emulator models of the current particle system in a reproducing kernel Hilbert space.
We here focus on modelling nonlinear covariance structure and gradients of the target.
The emulator's geometry is adaptively updated and subsequently used to inform local proposals.
Unlike in adaptive Markov chain Monte Carlo, continuous adaptation does not compromise convergence of the sampler.
KSMC combines the strengths of sequental Monte Carlo and kernel methods:
superior performance for multimodal targets and the ability to estimate model evidence  as compared to Markov chain Monte Carlo,
and the emulator's ability to represent targets that exhibit high degrees of nonlinearity.
As KSMC does not require access to target gradients, it is particularly applicable on targets whose gradients are unknown or prohibitively expensive.
We describe necessary tuning details and demonstrate the benefits of the the proposed methodology on a series of challenging synthetic and real-world examples.
\end{abstract}

\section{Introduction}

Monte Carlo methods for estimating integrals have become one of the main inference tools of statistics and machine learning over the last thirty years.
They are used to numerically approximate intractable integrals with respect to Bayesian posterior distributions.
Importantly, they also provide means to quantify uncertainty in the form of variance estimates, credible intervals and regions of high posterior density.
The most widely adopted Monte Carlo method is Markov Chain Monte Carlo (MCMC), which constructs a Markov chain that admits the desired target as its stationary distribution;
MCMC generates approximate samples from the target when the chain is run sufficiently long.
Poorly tuned MCMC samplers may need to run `burn in' for a very long time before reaching its equilibrium distribution, and successive samples may be highly correlated.

In contrast, sequential Monte Carlo (SMC) methods are based on iterative importance sampling, and have traditionally been applied to inference in filtering problems with a sequence of time-varying target distributions \citep{Doucet2001}, e.g.~in state-space models, where each intermediate distribution is typically defined on a successively larger latent space.
In this paper, we focus on static SMC methods, which recently have generated increasing interest as an alternative to MCMC for Bayesian inference on a single target distribution \citep{Cappe2008,Chopin2002,DelMoral2006,Fearnhead2013}.
Static SMC frames inference over a fixed target distribution as a sequential problem by defining an artificial series of incremental targets. This can be done by tempering the target density \citep{DelMoral2006},
by including data points sequentially \citep{Chopin2002}, or by targeting the full density at every iteration. The latter is a special case known as population Monte Carlo \citep[PMC,][]{Cappe2004}.

Kernel methods have recently been employed to construct efficient adaptive MCMC algorithms:
via modelling a Markov chain trajectory in a reproducing kernel Hilbert space (RKHS) and using geometry therein, it is possible to significantly improve mixing on target distributions with nonlinear interactions between components.
Covariance in the RKHS can be used to construct an adaptive random walk scheme, kernel adaptive Metropolis Hastings (KAMH), with proposals that are locally aligned with the target density \citep{SejStrGarAndGre14}.
Gradients of exponential families in the RKHS can be used to construct kernel Hamiltonian Monte Carlo (KHMC), an algorithm that behaves similar to Hamiltonian Monte Carlo (HMC) but without requiring access to gradient information \citep{strathmann2015gradient}.
Both KAMH and KHMC fall back to a random walk in yet unexplored regions, inheriting convergence properties such as geometric ergodicity on log-concave targets \citep[c.f. Proposition 3 in][]{strathmann2015gradient}.

In this paper, we develop a framework for \emph{kernel sequential Monte Carlo (KSMC)} for sampling from static models.
Similarly to the previous work in adaptive MCMC \citep{SejStrGarAndGre14, strathmann2015gradient}, KSMC represents the (weighted) particle system of SMC algorithms in a RKHS.
The learned geometry of the corresponding `emulator' model is used to construct proposal distributions for both MCMC rejuvenation and importance sampling steps inside SMC.

We apply this framework to two existing SMC algorithms, 
combining the strengths of SMC with those of kernel adaptive MCMC.
Firstly, we introduce \emph{kernel adaptive sequential Monte Carlo (KASMC)}, where the global covariance estimate in the adaptive SMC sampler  \citep[ASMC,][]{Fearnhead2013} is replaced by a kernel-informed local covariance \citep{SejStrGarAndGre14}.
Similar to ASMC, KASMC's proposals start as a standard random walk and then smoothly transition to taking locally aligned steps.
As a result, sampling efficiency can be significantly improved over ASMC.
Secondly, we use an infinite dimensional exponential family model \citep{SriFukKumGreHyv14} to estimate target gradients as in \citet{strathmann2015gradient}.
This results in \emph{kernel gradient importance sampling (KGRIS)}, a gradient-free version of gradient importance sampling (GRIS) \citep{schuster2015gradient}.
KGRIS is a novel adaptation of kernel gradient estimation ideas for constructing Langevin diffusions, and inherits their sampling efficiency compared to random walks.
Our contribution includes crucial implementation details, such as Rao-Blackwelisation, stratification, and tuning of the presented algorithms.

Unlike for Langevin diffusions or Hamiltonian dynamics, our framework does not require gradients or higher-order information of the target. 
Consequently, the KSMC framework is particularly useful in combination with importance sampling frameworks 
such as SMC${}^2$ \citep{chopin2013smc2} and IS${}^2$ \citep{Tran2013} 
for sampling from doubly intractable targets, where gradient information is unavailable.

We finally argue that (adaptive) SMC is a more natural framework for employing RKHS-based representations.
Adaptive MCMC samplers require a vanishing adaptation schedule in order to ensure convergence to the correct target \citep{Rosenthal2011},
creating a difficult to tune exploration-exploitation trade-off with limited principled guidance on selecting such adaptation schedules.
In contrast, SMC proposals can continuously be adapted and the choice of an adaptation schedule is thus entirely circumvented. An easy to use Python package implementing the proposed methods is available under an open source licence.\footnote{Source code available at \url{https://github.com/ingmarschuster/kameleon_rks}}

\begin{algorithm*}[tb]
	\caption{Sequential Monte Carlo for Static Models}
	\begin{algorithmic}
		\label{algo:smc}
		\STATE {\bfseries Input:} Sequence of target densities $\targd_0,\dots,\targd_T$ (where $\targd_T = \targd$), size of particle system $\psize$ 
		\STATE {\bfseries Output:} sets $\SaS_1,\dots,\SaS_T$ and $\Wset_1,\dots,\Wset_T$ of samples and accompanying weights
		\STATE Initialise $\SaS_0$ to $\psize$ samples from $\targd_0$, and $\Wset_0$ to equal weights $1/\psize$
		\FOR{$t=1$ through $t=T$}
		\STATE $\RWset_t = \{ {\iw_{t-1}^i \targd_t(\smp_{t-1}^{i})}/{\targd_{t-1}(\smp_{t-1}^{i})} \}_{i = 1}^\psize$ 
		\STATE construct $\RSaS_t$ by re-sampling $( \SaS_{t-1}, \RWset_t)$, resulting in $\psize$ copies of samples in $\SaS_{t-1}$ 
		\STATE construct or update proposal $\propd_t$
		\IF{using an MH transition kernel}
		\STATE Set $\SaS_t$ to \\~~~$\{ \smp_t^i \sim \textrm{MH kernel with proposal }\propd_t(\cdot|\Rsmp_t^i) \}_{i =1}^\psize$
		\STATE $\Wset_t = \{ 1/\psize \}_{i=1}^\psize$
		\ELSE 
		\STATE Set $\SaS_t$ to $\psize$ samples from\\ ~~~$\propd_t^\mathrm{Mixt}(\cdot) = \frac{1}{\psize} \sum_{i=1}^\psize \propd_t(\cdot|\Rsmp_t^i)$
		\STATE $\Wset_t = \{ \targd_t(\smp_{t,i})/\propd_t^\mathrm{Mixt}(\smp_{t,i}) \}_{ i = 1}^\psize$
		\ENDIF
		\ENDFOR
		\STATE return $\SaS_1,\dots,\SaS_T$ and $\Wset_1,\dots,\Wset_T$
	\end{algorithmic}
	
\end{algorithm*}

\section{Background}
\label{sec:ismc}

Sequential Monte Carlo algorithms \citep{DelMoral2006,Doucet2009} approximate a target density $\pi$ 
by iteratively targeting a sequence of incremental densities $\targd_0, \dots, \targd_T$, with $\targd_T = \targd$.
These incremental densities are typically defined such that the initial density $\targd_0$ is easy to sample from (e.g.\ the prior in a Bayesian model).  
Consecutive distributions $\targd_t, \targd_{t+1}$ are `close', in the sense that drawing samples from $\targd_{t+1}$ given samples from $\targd_t$
is easier than drawing samples from $\targd_{t+1}$ directly. 
At each stage $t$, we approximate the target density $\targd_t$ with a set of $\psize$ samples $\SaS_t = \{\smp_t^{i} \}_{i=1}^\psize$
with associated importance weights $\Wset_t = \{ \iw_t^{i} \}_{i=1}^\psize$, with
\begin{align}
\hat \targd_t(\smp) = \sum_{i=1}^\psize \iw_t^i \delta_{\smp_t^i}(\smp)
\label{eq:approx-is}
\end{align}
where $\delta_{\smp_t^i}$ is a Dirac point mass on $\smp_t^i$.
In contrast to SMC as applied to state space models, in a static SMC setting each target density $\targd_t$ is defined on the same space $\mathcal{X}$.

We initialise the algorithm by sampling an initial set of $\psize$ samples $\SaS_0$ 
from the initial density $\propd_0$,
with equal importance weights $\nicefrac{1}{\psize}$.
For each subsequent $t = 1,\dots,T$, given a particle set $(\SaS_{t-1}, \Wset_{t-1})$ approximating $\targd_{t-1}$,
we construct a new particle set which approximates $\targd_t$.
This is a three-step process, summarised in Algorithm \ref{algo:smc}.
First, we re-weight each particle relative to the new target density, setting
\begin{align*}
\Riw_t^i &= \iw_{t-1}^i \frac{\pi_t(\smp_{t-1}^i)}{\pi_{t-1}(\smp_{t-1}^i)}.
\end{align*}
Weighting the points in $\SaS_{t-1}$ by $\{ \Riw_t^i \}_{i=1}^\psize$ yields an approximation to $\pi_t$ 
in the same manner as in \eqref{eq:approx-is} ---
the new importance weights correct for the change from $\targd_{t-1}$ to $\targd_{t}$.

Static SMC then applies re-sampling, 
constructing an equally-weighted set of particles $\RSaS_t = \{ \Rsmp_t^i \}_{i=1}^\psize$ 
by sampling with replacement from $\SaS_{t-1}$ with weights proportional to $\Riw_t^i$, \citep{Douc2005}.
Together, these samples form an approximation to $\targd_t$,
where values from $\SaS_{t-1}$ with high weight under $\targd_{t}$ have been duplicated
and those with low weight under $\targd_{t-1}$ have been discarded.
This duplication of values, however, can lead to a sample impoverishment problem: 
many of the re-sampled values $\Rsmp_t^i$ may have identical values.
This can be avoided by applying a so-called {\em rejuvenation} step after re-sampling \citep{Chopin2002},
constructing an overall approximation $( \SaS_t, \Wset_t )$ to $\targd_t$ with a diverse set of values of $\smp_t^i$.

The rejuvenation step consists of a proposal $\propd_t(\SaS_t | \RSaS_t)$.
We here consider two ways of incorporating such a proposal. 
One traditional option is to use a Markov density $\propd_t$ as a proposal in a Metropolis-Hastings (MH) kernel which leaves $\pi_t$ invariant: 
For each $\Rsmp_t^i$ in $\RSaS_t$, we propose a new value $\smp_t^i$ from $\propd_t(\smp_t^i | \Rsmp_t^i)$ and accept it according to a standard MH acceptance ratio targeting $\pi_t$.
In this case, each importance weight in $\Wset_t$ will be identically $\nicefrac{1}{\psize}$.

An alternative is to consider the mixture proposal of all such Markov densities $\propd_t$ as an importance sampling proposal over $\pi_t$,
a common approach in PMC.
We can define 
\begin{align*}
\propd_t^\mathrm{Mixt}(\smp_t) = \frac{1}{\psize} \sum_{i=1}^\psize \propd_t(\smp_t | \Rsmp_t^i),
\end{align*}
and draw $\psize$ samples $\smp_t$ from $\propd_t^\mathrm{Mixt}$ to generate $\SaS_t$.
Now we set  importance weights in $\Wset_t$
to $\iw_t^i = \targd_t(\smp_t^i) / \propd_t^\mathrm{Mixt}(\smp_t^i)$ for $i = 1,\dots,\psize$.



\subsection{Existing SMC algorithms}

In SMC algorithms, we are free in choosing a proposal $q_t$. In contrast to MCMC, it may be directly informed by the previous samples
$\SaS_{t-1}$ and their weights $\Wset_{t-1}$.
The following two existing SMC algorithms are examples that we will extend to kernel-based alternatives.

\paragraph{Adaptive SMC}

The adaptive SMC sampler (ASMC) studied by \citet{Fearnhead2013} is based on continuously estimating the global covariance $\Sigma_t$ of $\targd_t$, and updating a scaling parameter $\nu^2$. 
This is done from the re-weighted particle system, which is subsequently moved through a Markov kernel.
The proposal distribution used within the MH kernel at point $\smp$ in Algorithm \ref{algo:smc} is $\propd_t(\cdot|\smp) = \DNorm(\cdot| \smp, \nu^2 \Sigma_t+\gamma^2I)$.

\paragraph{Gradient importance sampling}
In addition to using the estimated covariance $\Sigma_t$ of $\targd$ as in ASMC, gradient importance sampling  \citep[GRIS,][]{schuster2015gradient} incorporates a drift term based on the log target gradient. 
For target gradient  ${\nabla \log \targd}$ and previous sample $\smp$, the proposal distribution in Algorithm \ref{algo:smc} is $\propd_t(\cdot) = \DNorm(\cdot|\smp + D(\nabla \log {\targd}(\smp)), \nu^2 \Sigma_t)$, for each individual particle $\smp$ in the current (unweighted) particle set. A typical choice for the drift function is $D(y) = \delta y$ with $0<\delta<1$. 
Rather than incorporating a MH step, the updated values are importance weighted  --- 
GRIS is a population Monte Carlo (PMC) algorithm.
In numerical experiments, GRIS compares favourably to its closest MCMC relatives like the adaptive MALTA algorithm and adaptive Metropolis \citep{schuster2015gradient}.

\subsection{Kernel adaptive MCMC proposals}

The previously described SMC algorithms are based on target covariance and gradients. 
We now review how these quantities were previously modelled using kernel methods in the context of MCMC. 
Note that any form of adaptation in MCMC requires care in order to preserve  ergodicity of the resulting Markov chain, and 
some form of vanishing adaptation is needed \citep{Andrieu2008,Rosenthal2011}.
This can be achieved e.g. by updating the proposal family with vanishing probability \citep{SejStrGarAndGre14, strathmann2015gradient}.

\label{sec:kernel_mcmc}
\paragraph{Covariance emulator}
\citet{SejStrGarAndGre14} introduced a kernel covariance emulator as a method for adapting the proposal distribution in a Metropolis-Hastings MCMC algorithm,
based on the history of the Markov chain $\SaS=\{\smp_1, \smp_2, \dots\}$.
The idea is to represent covariance of the target as an empirical Gaussian measure with mean
$\mu_\SaS:=\frac{1}{\vert \SaS\vert}\sum_{\smp \in \SaS}\PDK(\smp, \cdot)$
and covariance $\frac{1}{\vert \SaS\vert}\sum_{\smp \in \SaS}\PDK(\smp, \cdot)\otimes  \PDK(\smp, \cdot) - \mu_\SaS \otimes \mu_\SaS$
in a RKHS with kernel $\PDK$.
This measure 
can be sampled from exactly, and it is possible to (approximately) map samples back to the original space.

\citet{SejStrGarAndGre14} showed that it is possible to integrate out the RKHS proposal
analytically, which elegantly results in a \emph{closed form} Gaussian proposal density in the input space.
For a Gaussian kernel, the proposal at particle ${\smp}_j$ locally aligns to the structure of the posterior at ${\smp}_j$, and is given by 
\begin{equation*}
\propd_{\text{KAMH}}(\cdot|{\smp}_j)=\DNorm(\cdot|{\smp}_j, \gamma^2I+\nu^2 M_{\SaS,{\smp}_j} C M_{\SaS,{\smp}_j}^\trans),
\end{equation*}
where $C=I-\frac{1}{n}11^\trans$ is a centering matrix and $M_{\SaS,{\smp}_j}$ collects kernel gradients with respect to all particles,
$$M_{\SaS,{\smp}_j} =2[\nabla_x\PDK(x,{\smp}_{1})|_{x={\smp}_j},...,\nabla_x\PDK(x,{\smp}_{\psize})|_{x={\smp}_j}].$$

Additional exploration noise with variance $\gamma^2$ avoids that the proposal collapses in unexplored regions of the input space.


\paragraph{Gradient emulator}
To overcome random walk behaviour of KAMH, \citet{strathmann2015gradient} constructed an algorithm that adaptively learns the gradient structure of the Markov chain history, and mimics Hamiltonian dynamics using the learned gradients.
This is done by fitting an un-normalised infinite dimensional exponential family model with density function $\exp(\langle f, k(x,\cdot)\rangle_\mathcal{H} -A(f))$. Here, $\langle f, k(x,\cdot)\rangle_\mathcal{H}=f(x)$ is the inner product between natural parameters $f$ and sufficient statistics $k(x,\cdot)$ in a RKHS $\mathcal{H}$, and $A(f)$ is the (intractable) log-partition function.
Remarkably, it is possible to efficiently estimate $f$ via minimising the expected $L^2$ error of $\nabla_x f(x)$ without dealing with $A(f)$.
Combining this model with a further approximation, based on random basis functions \citep[KMC finite;][]{strathmann2015gradient}, allows for efficient on-line updates of the emulator.
Similar to Hamiltonian Monte Carlo, the resulting KHMC algorithm offers substantial improvements over random walks. Tt does so, however, \emph{without} requiring gradient information of the target.
This allows application to intractable likelihood models, where we cannot evaluate the target densities $\pi_t$ even up to a normalizing constant, and gradients are similarly unavailable.

\section{Kernel sequential Monte Carlo}
\label{sec:kam_smc}

We now develop a kernel sequential Monte Carlo framework. 
KSMC is based on combining classical adaptive SMC with the emulator based proposals of kernel adaptive MCMC.
In general, once a kernel emulator is fitted to past particle systems, we can use it in either of two ways: as proposals for MH rejuvenation steps inside SMC or as importance densities in PMC.

\paragraph{Key contributions.}
Our main contribution is to combine several yet unconnected pieces of literature into a novel framework that performs favourably compared to its individual parts: adaptive SMC proposals, SMC for intractable likelihoods, and kernel emulators for efficient proposals.
This combination is simple yet very natural: As compared to (kernel) adaptive MCMC, the KSMC framework (i) circumvents the need for vanishing adaptation, (ii) can represent multimodality, (iii) allows to estimate model evidence in a straight-forward manner.
On the other hand, as compared to plain adaptive SMC and PMC, the use of kernel emulators (iv) leads to faster convergence for nonlinear targets.

We present two novel algorithms, KASMC and KGRIS, both of which are weighted and kernelised generalisations of existing kernel MCMC and SMC respectively. 
These modifications can lead to significant mixing improvements in practice.
Our contribution furthermore includes variance reduction techniques that are critical in practice. In particular, na\"ive implementations can suffer from high variance induced by simplifications. As this results in lower quality emulators, too high variance would be self-reinforcing 
and is to be strictly avoided.

\subsection{Kernel adaptive rejuvenation: KASMC}
\label{sec:kam_rejuv}

We can use both kernel emulators for the rejuvenation step of SMC. 
More specifically, at time-step $t+1$, we target distribution $\targd_{t+1}$, based on a particle system approximating $\targd_{t}$.
After re-weighting, the new system $\{(\iw_{t+1,i}, \smp_{t+1,i})\}_{i=1}^\psize$ is a weighted approximation to $\targd_{t+1}$.
We here focus on the nonlinear covariance emulator which can be either fitted using the equally-weighted re-sampled values $\RSaS_t$, or the original particle set with weights $\RWset_t$.
The proposal distribution for Algorithm \ref{algo:smc} at $\smp$ then is exactly  $\propd_{\text{KAMH}}$. As in KAMH, this results in covariance matrices for Gaussian proposals which locally align with the target \citep{SejStrGarAndGre14}, now taking the SMC particle weights into account.
The resulting kernel adaptive SMC sampler (KASMC) inherits KAMH's ability to explore non-linear targets more efficiently than proposals based on estimating global covariance structure such as in \citet{Fearnhead2013} and \citet{Haario2001}.
Figure \ref{fig:illustration_and_banana} (left) shows a simple illustration of a global (ASMC) and local proposal distribution (KASMC).
Compared to previous work on kernel induced local covariance matrices for MCMC \citep{SejStrGarAndGre14}, we implement a random features approximation in order to enable computationally efficient updates with  information gained from new samples \citep{Rahimi2007}.

\subsection{Kernel induced importance densities: KGRIS}

Another way to use kernel-based emulators is for generating proposals which are corrected by importance sampling, i.e.\ in PMC.
In our second approach, a kernel emulator is fitted to weighted particles, which were previously corrected via importance weights.
As an example, we here use the kernel gradient emulator by \citet{strathmann2015gradient}, in its finite dimensional approximation (KMC finite), c.f.\ \citep[Proposition 2]{strathmann2015gradient}.

The log density of the approximate estimator takes the simple form $f(x)=\theta^\top \phi_x$, where $\phi_x\in\mathbb{R}^m$ is an embedding of $x$ into an $m$-dimensional feature space, and $\theta\in\mathbb{R}^m$ is estimated by $\hat{\theta}=C^{-1}b$ from samples $x$.
Given a weighted particle system $\{(\iw_{t,i}, \smp_{t,i})\}_{i=1}^\psize$, then $b,C$ are weighted averages of the form
\begin{align*}
b:=-\frac{1}{\sum_{i=1}^\psize \iw_{t,i}}\sum_{i=1}^\psize\iw_{t,i}\sum_{\ell=1}^{d}\ddot{\phi}_{x}^{\ell},\\ C:=\frac{1}{\sum_{i=1}^\psize \iw_{t,i}}\sum_{i=1}^\psize\iw_{t,i}\sum_{\ell=1}^{d}\dot{\phi}_{x}^{\ell}\left(\dot{\phi}_{x}^{\ell}\right)^{\top},
\end{align*}
with element-wise derivatives $\dot{\phi}_{x}^{\ell}:=\frac{\partial}{\partial x_{\ell}}\phi_{x}$
and $\ddot{\phi}_{x}^{\ell}:=\frac{\partial^{2}}{\partial x_{\ell}^{2}}\phi_{x}$. Note that the estimator can be updated in an online fashion once the particle system changes. 
Rather than simulating Hamiltonian dynamics to generate a proposal, we here take single gradient steps, i.e.\ the Markov density at in Algorithm~\ref{algo:smc} at $\smp$ is $\propd_t(\cdot|\smp) = \DNorm(\cdot|\smp + \delta \nabla f(\smp), \nu^2 \Sigma_t)$
for some parameters $\delta > 0, \nu^2 > 0$.
This keeps the risk of divergence due to wrongly estimated gradients low.
We arrive at kernel GRIS, a gradient-free variant of GRIS \citep{schuster2015gradient}.

\subsection{Controlling emulator variance in PMC}
PMC is somewhat sensitive to badly scaled proposals, as these are not rejected as in a Metropolis-Hastings step.
In particular for gradient emulators used within PMC, variance reduction is important to avoid numerical divergence.
The original PMC paper introduces re-sampling in order to deal with un-weighted instead of weighted samples \citep{Cappe2004}, though at the cost of an increased variance.
While some approaches avoid re-sampling altogether \citep{Cappe2008}, 
we consider re-sampling here as a way to obtain a set of locations $\{ \Rsmp_t^i \}_{i=1}^\psize$ for our Markov proposal components of the mixture 
$\propd_t^\mathrm{Mixt}(\cdot) = \frac{1}{\psize} \sum_{i=1}^\psize \propd_t(\cdot|\Rsmp_t^i)$, due to better behaving variance in high dimension.
With re-sampling, Monte Carlo variance only grows as $O(\targdim)$ rather than $O(\exp(\targdim))$ without re-sampling, where $\targdim$ is the dimensionality \citep{Doucet2009}.

Given a re-sampled number of $\psize$ particles and the updated emulator $\propd_t$, we simulate from the mixture distribution $\propd^\mathrm{Mixt}_t$ with stratification, i.e.\ we draw exactly one sample from each of the equally weighted mixture components. Another view of this scheme is to draw a single realisation from $\propd_t(\cdot|\Rsmp_t^i)$ for all $i=1,\dots,\psize$ and Rao-Blackwellise. 
Finally, we can view the scheme as an instance of the deterministic mixture idea  \citep{Elvira2015}.
Without this technique, i.e.\ using weights $\targd(\cdot)/\propd_t(\cdot|\Rsmp_t^i)$, variance might grow catastrophically large, as too high variance can be self-reinforcing by resulting in emulators of low quality.

\section{Evaluation}
\label{sec:eval}
We empirically evaluate performance of KASMC on a simple non-linear target,
on a multi-modal sensor network localisation problem,
and in estimating Bayesian model evidence in a model with an intractable likelihood on a real-world dataset.
The final experiment uses a challenging stochastic volatility model with S\&P $500$ data from \citet{chopin2013smc2} to evaluate KGRIS.

For the KASMC experiments on static target distributions, 
a sequence of incremental target densities can be defined using a geometric bridge
with $\targd_t \propto \targd_0^{1-\rho_t} \targd^{\rho_t}$ for some initial distribution $\targd_0$, where $(\rho_t)_{t=1}^T$ is an increasing sequence satisfying $\rho_T = 1$.
The bandwidth parameter of the kernel emulator models is set to the median distance between particles \cite{gretton2012kernel}.

We also note these algorithms have a free scaling parameter $\nu^2$, which we would like to adapt online. 
To accomplish parameter tuning, we use the standard framework of stochastic approximation for tuning MCMC kernels \cite{Andrieu2008}, i.e.\ tuning acceptance rate $\alpha_t$  towards an asymptotically optimal acceptance rate  $\alpha_\textrm{opt}=0.234$ for random walk proposals \cite{Rosenthal2011}.
After the MCMC rejuvenation step, a Rao-Blackwellised estimate $\hat\alpha_t$ of expected acceptance probability is available by simply averaging the acceptance probabilities for all MH proposals.
	Then, set
	$\nu^2_{t+1}=\nu^2_t + \lambda_t(\hat\alpha_t-\alpha_\textrm{opt})$
	for some non-increasing sequence $\lambda_1,\dots,\lambda_T$.
	This strategy of approximating optimal scaling assumes that consecutive targets are close enough so that the acceptance rate when using $\nu^2_t$ to target $\targd_t$ provides information about the expected acceptance rate when using $\nu^2_t$ with target $\targd_{t+1}$. This is discussed further in the supplemental material.
	
\subsection{KASMC: Improved convergence on synthetic nonlinear target}
\label{sec:eval:banana}

\begin{figure*}[t]
	\begin{minipage}{0.5\textwidth}
		\includegraphics[height=1in]{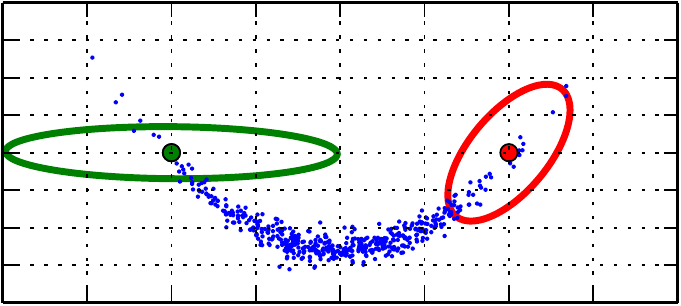}
		\vspace{0.15in}
	\end{minipage}
	\hfill
	\begin{minipage}{0.5\textwidth}
		\includegraphics[width=\textwidth]{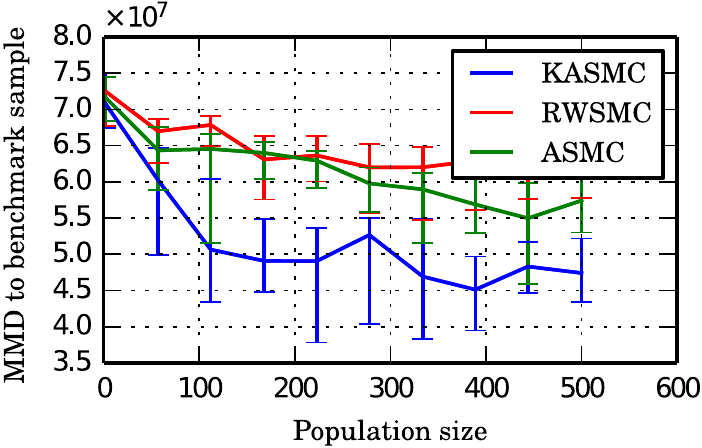}
	\end{minipage}
	\caption{\textbf{Left:} Proposal distributions around one of many particles (blue) for each KASMC (red) and ASMC (green).
		KASMC proposals locally align to the target density while ASMC's global covariance estimate might result in poor MH rejuvenation moves. \textbf{Right:} Improved convergence of all mixed moments up to order 3 of KASMC compared to using SMC with static or adaptive Metropolis-Hastings steps. }
	\label{fig:illustration_and_banana}
\end{figure*}

We begin by studying  convergence of KASMC compared to existing algorithms on a simple benchmark example: the strongly twisted banana-shaped distribution in $D=8$ dimensions used in \citet{SejStrGarAndGre14}.
This distribution is a multivariate Gaussian with a non-linearly transformed second component,
defined as
	\[
	\mathcal{B}(y;b,v)=\mathcal{N}(y_{1};0,v)\mathcal{N}(y_{2};b(y_{1}^{2}-v),1)\prod_{j=3}^{\targdim}\mathcal{N}(y_{j};0,1).
	\]
We compare SMC algorithms using different rejuvenation MH steps: a static random walk Metropolis move (RWSMC) with fixed scaling  $\nu=2.38/\sqrt{\targdim}$, ASMC, and KASMC using a Gaussian RBF kernel. 
For the latter two algorithms, all particles are used to compute the proposal, and a fixed learning rate of $\lambda=0.1$ is chosen to adapt scale parameters.
Starting with particles from a multivariate Gaussian $\mathcal{N}(0,50^2)$, we use a geometric bridge that reaches the target $\mathcal{B}(y;b=0.1,v=100)$ in 20 steps.
We repeat the experiment over $30$ runs.
Figure \ref{fig:illustration_and_banana} (right) shows that KASMC achieves faster convergence of the first 3 moments, i.e. in MMD\footnote{The maximum mean discrepancy, here using a polynomial kernel of order 3, quantifies differences of all mixed moments up to order 3 of two independent sets of samples.} distance to a large benchmark sample. 

%
%

	\subsection{A multi-modal application: sensor network localisation}
	\label{sec:eval:sensor_networks}

	We next study performance of KASMC on a multi-modal target arising in a real-world application: inferring the locations of $S$ sensors within a network, as discussed in \cite{ihler2005nonparametric, lan2014wormhole}.
	We here focus on the static case: assume a number of stationary sensors that measure distance to each other in a $2$-dimensional space;
	a distance measurement is successful with a probability that decays exponentially in the squared distance, and the observation is missing otherwise.
	If distance is measured, it is corrupted by Gaussian noise.
	The posterior over the unknown sensor locations forms an extremely constrained non-linear and multi-modal distribution induced by the spatial set-up.
	
	Assume $S$ sensors with unknown locations $\{x_i\}_{i=1}^S\subseteq \mathbb{R}^2$.
	Define an indicator variable $Z_{i,j}\in\{0,1\}$ for the distance $Y_{ij}\in\mathbb{R}^+$ between a pair of sensors $(x_i, x_j )$ being either observed ($Z_{i,j}=1$) or not ($Z_{i,j}=0$), according to $$Z_{i,j}\sim\texttt{Binom}\left(1,\exp\left(-\frac {\Vert x_i - x_j\Vert_2^2}{2R^2}\right)\right).$$ If the
	distance is observed, then $Y_{ij}$ is corrupted by Gaussian noise, i.e. $$Y_{i,j}|Z_{i,j}=1\sim\mathcal{N}\left(\Vert x_i-x_j\Vert,\sigma^2\right),$$ and $Y_{i,j}=0$ otherwise.
	
	Previously, \cite{ihler2005nonparametric} focussed on MAP estimation of the sensor locations, and \cite{lan2014wormhole} focussed on a well-conditioned case ($S=8$ sensors and $B=3$ base sensors with known locations) that results in almost no ambiguity in the posterior.
	We argue that Bayesian quantification of uncertainty is more important for cases where noise and missing measurements \emph{does not} allow to reconstruct the sensor locations exactly.
	We therefore reuse the dataset from \cite{lan2014wormhole} ($R=0.3$, $\sigma^2=0.02$)\footnote{Downloaded from \url{http://www.ics.uci.edu/~slan/lanzi/CODES_files/WHMC-code.zip} on 8/Oct/2015.}, but only use the first $S=3$ locations/observations.
	In order to encourage ambiguities in the localisation task, we only use the first $2$ base sensors of \cite{lan2014wormhole} with known locations that each do observe distances to the $S$ unknown sensors but not of each other.
	Unlike \cite{lan2014wormhole}, we use a Gaussian prior $\mathcal{N}(\mathbf{0.5}, I)$ to avoid the posterior being situated in a bounded domain.

\begin{figure*}[t]
		\centering
		\includegraphics[width=0.5\textwidth]{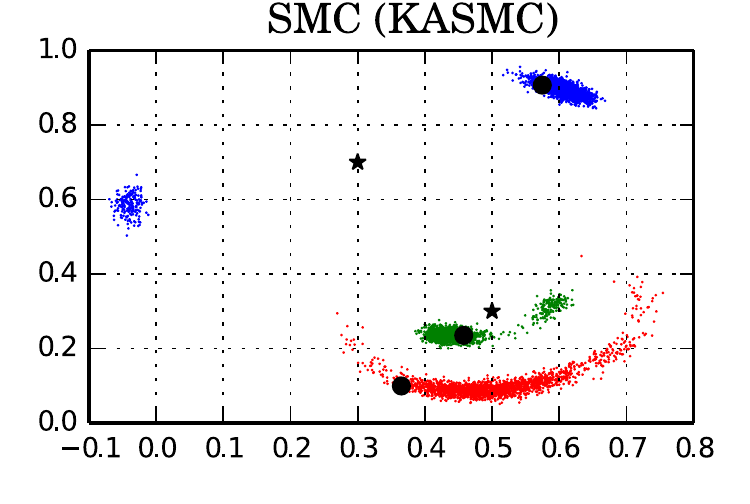}
		\hfill
		\includegraphics[width=0.5\textwidth]{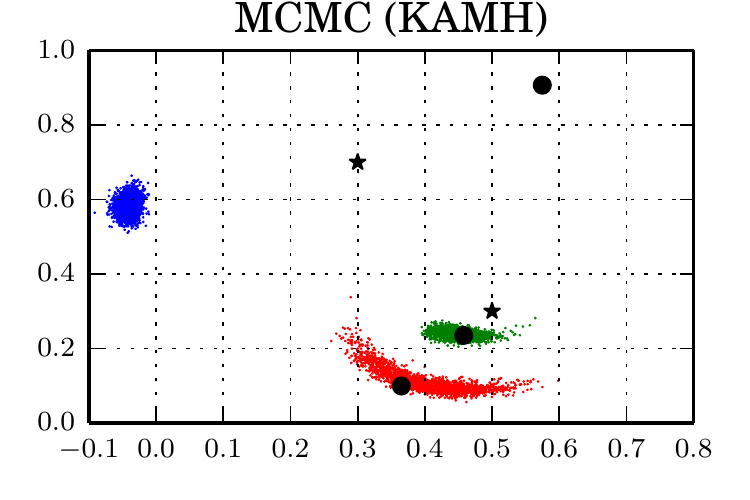}
		\caption{Posterior samples of unknown sensor locations (in color) by kernel-based SMC and MCMC on the sensors dataset.
			The set-up of the true sensor locations (black dots) and base sensors (black stars) causes uncertainty in the posterior.
			SMC recovers all modes while MCMC does not.
			The posterior has a clear non-linear structure.}
		\label{fig:eval:smc_mcmc_sensors}
	\end{figure*}

	Figure \ref{fig:eval:smc_mcmc_sensors} shows the marginalised posterior for one run each of KASMC (SMC) and KAMH (MCMC), where we matched the number of likelihood evaluations (500,000).
	We run KASMC using $10,000$ particles and a bridge length of $50$, and MCMC-KAMH for $50 \times 10,000$ iterations of which we discard half as burn-in; both were initialized with samples from the prior.
	Tuning parameters $\nu^2$ are set using a diminishing adaptation schedule $\lambda_t=1/\sqrt{t}$ for KAMH and a fixed learning rate $\lambda_t=1$ for KASMC.
	MCMC is not able to traverse between the multiple modes and interpretations of the data, in contrast to SMC.

	In order to compare ASMC to KASMC, we created a benchmark sample via running 100 standard MCMC chains (randomly initialised to cover all modes) each for 50000 iterations, discarding half the samples as burn-in, and randomly down-sampling to a size of 100. 
	We then compute the empirical MMD distance to the output of the individual algorithms, averaged over $10$ runs.
	For the chosen number of sensors, ASMC and KASMC perform similarly.
	With less sensors, i.e.\ more ambiguity, KASMC produces samples with both less MMD distance from a benchmark sample and less variance.
	For example, for a set-up with $S=2$ and 1000 particles, we get a MMD distance to a benchmark sample of $0.76 \pm 0.4$ for KASMC and $0.94 \pm 0.7$ for ASMC.


\subsection{KASMC: evidence estimation in Gaussian process classification}
\label{exp:gp}
Following \citet{SejStrGarAndGre14}, we consider Bayesian classification on the UCI Glass dataset, discriminating window glass from non-window glass, using a Gaussian process (GP).
It was found that the induced posterior is indeed non-linear \citep{strathmann2015gradient, SejStrGarAndGre14}. 
In \citet{SejStrGarAndGre14}, samples from the marginal posterior over GP hyper-parameters were simulated (the GP latent variables integrated out).
We emphasise a different point here: KSMC's ability to estimate the model evidence as compared to KAMH, and its faster convergence compared to ASMC.

Consider the joint distribution of latent variables $\mathbf{f}$,
labels $\mathbf{y}$ (with covariate matrix \textbf{$X$}), and hyper-parameters
$\theta$, given by
\[
p(\mathbf{f},\mathbf{y},\theta)=p(\theta)p(\mathbf{f}|\theta)p(\mathbf{y}|\mathbf{f}),
\]
where $\mathbf{f}|\theta\sim{\cal N}(0,\mathcal{K}_{\theta})$, with
$\mathcal{K}_{\theta}$ modelling the covariance between latent variables
evaluated at the input covariates. 
Consider the binary logistic classifier, i.e.\ $p(y_{i}|f_{i})=\frac{1}{1-\exp(-y_{i}f_{i})}$
where $y_{i}\in\{-1,1\}$.
In order to perform Bayesian model selection (i.e.\ comparing different covariance functions), we need to estimate the model evidence of the marginal posterior given the hyper-parameters. 
Here, the marginal likelihood $p(\mathbf{y}|\theta)$ is intractable for non-Gaussian likelihoods $p(\mathbf{y}|\mathbf{f})$.
We estimate model evidence for the GP classifier equipped with a standard Gaussian Automatic Relevance Determination (ARD) covariance kernel;
an unbiased estimate can be obtained using importance sampling
\begin{equation}
\hat{p}(\mathbf{y}|\theta):=\frac{1}{n_{\textrm{imp}}}\sum_{i=1}^{n_{\textrm{imp}}}p(\mathbf{y}|\mathbf{f}^{(i)})\frac{p(\mathbf{f}^{(i)}|\theta)}{q(\mathbf{f}^{(i)}|\theta)},
\end{equation}
where $\left\{ \mathbf{f}^{(i)}\right\} _{i=1}^{n_{\textrm{imp}}}\sim q(\mathbf{f}|\theta)$
are $n_{\textrm{imp}}$ importance samples, e.g.\ from a Laplace approximation of $p(\mathbf{f}|\mathbf{y},\theta)$.
We here do not tune the number of 'inner' importance samples, but follow \cite{SejStrGarAndGre14} and use $n_\text{imp}=100$.

Figure \ref{fig:eval:gp} shows that evidence estimates of KASMC exhibit less variance than those of ASMC.
The ground truth model evidence was established via running 20 SMC instances using $\psize=1000$ particles and a bridge length of 30, and averaging their evidence estimates.
 The experiment is performed 50 times, using $\psize=100$ particles and a bridge length of 20, starting from he prior on the log hyper-parameters $\targd_0 = p(\theta_d) \equiv \mathcal{N}(0,5^2)$.
The learning rate is constant $\lambda_t=1$, and adaptation is towards an acceptance rate of $0.23$.

\subsection{KGRIS: stochasitic volatility model with intractable likelihood}
\label{exp:stoch_vol}

\begin{figure}[t]
\centering
\begin{minipage}{.42\textwidth}
  \centering
	\includegraphics[width=\textwidth]{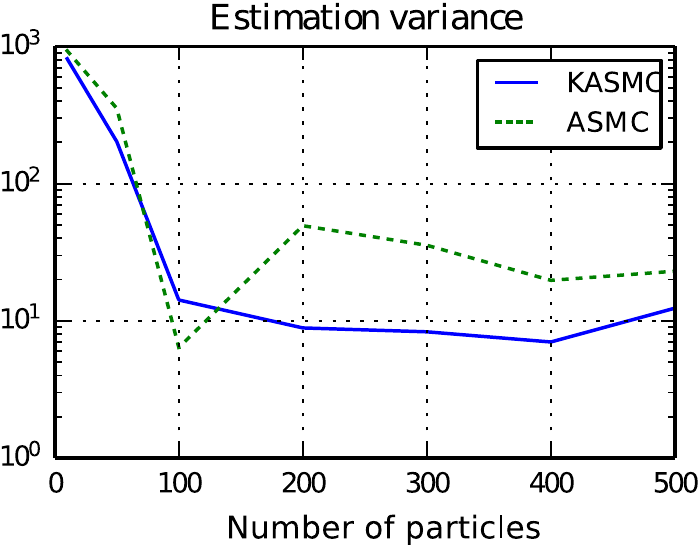}
	\captionof{figure}{Estimating model evidence of a GP using the IS$^2$ framework. The plot shows the MC variance over 50 runs as a function of the size of the particle system.}
	\label{fig:eval:gp}
\end{minipage}%
\hfill
\begin{minipage}{.5\textwidth}
  \centering
	\includegraphics[width=\textwidth]{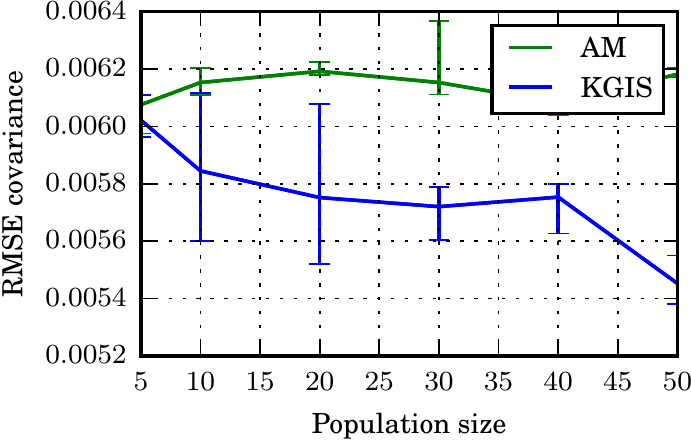}
	\captionof{figure}{Convergence of RMSE for estimating all elements of the posterior covariance matrix of the stochastic volatility model.
	}
	\label{fig:eval:sv}
	
\end{minipage}
\end{figure}

A particularly challenging class of Bayesian inverse problems are stochastic volatility models.
As time series models, they often involve high-dimensional nuisance variables, which usually cannot be integrated out analytically.
Furthermore, risk management necessitates to account for parameter and model uncertainty, and models have to capture the non-linearities in the data \citep{chopin2013smc2}.
We here concentrate on the prediction of daily volatility of asset prices, reusing the model and dataset studied by \citet{chopin2013smc2} to evaluate KGRIS.
Due to the lack of analytically available gradients for this model, we compare two gradient free PMC versions: KGRIS and a random walk PMC with global covariance adaptation in the style of \citet{Haario2001}.

Let $s_t$ be the value of some financial asset on day $t$, then $y_t = 10^{(5/2)}\log(s_t/s_{t-1})$ is called the log-returns (upscaling for numerical reasons).
We model the observed log-return $y_t$ as dependent on a latent $v_t$ by the observation equation 
$$y_t = \mu + \beta v_t + \sqrt{v_t}\epsilon_t$$ 
for $t \leq 1$.
Here $\epsilon_t$ is a sequence of i.i.d.\ standard Gaussian errors and $v_t$ is assumed to be a stationary stochastic process known as the \emph{actual volatility}.
\citet{chopin2013smc2} develop a hierarchical model for $v_t$ based on the idea of analytically integrating a continuous time volatility model over daily intervals \citep[for details see][]{chopin2013smc2}.
Using this construction, the (discrete time) $v_t$ is parameterised by stationary mean $\xi$ and variance $\omega^2$ of the so called \emph{spot volatility} and the exponential decay $\lambda$ of its auto-correlation.
This results in the following model for the actual volatility $v_t$:
\begin{gather*} k \sim \textrm{Pois}(\lambda\xi^2/\omega^2), c_{1:k} \sim \textrm{U}(t,t+1), e_{1:k} \sim \textrm{Exp}(\xi/\omega^2)\\
z_{t+1} = z_t \exp({-\lambda})+\sum_{j=1}^{k}e_j\exp({-\lambda(t+1-c_j)})\\
v_{t+1} = \lambda^{-1} \bigg( z_t-z_{t+1}+\sum_{j=1}^{k}e_j \bigg), x_{t+1} = (v_{t+1},z_{t+1})^\trans
\end{gather*}
where $z_t$ is the discretely sampled spot volatility process and $(v_{t+1},z_{t+1})^\trans$ is the Markovian representation of the state process.
The variables $k, c_{1:k}$ and $e_{1:k}$ are generated independently for each time period. For $k=0$, the set $1:k$ is defined to be empty.
The dynamics imply $\Gamma(\xi^2/\omega^2, \xi^2/\omega^2)$ to be the stationary distribution for $z_t$, which is also used as the initial distribution on $z_0$.
The parameters of the model are $\param = (\mu, \beta, \xi,\omega^2,\lambda)$ and the likelihood is intractable.  \citet{chopin2013smc2} developed a sampler for $\param$ based on iterated batch importance sampling using nested SMC with pseudo-marginal MCMC moves for integrating out the $x_t$ and dubbed their approach SMC$^2$.

In our experiment, we use KGRIS proposals in a population Monte Carlo setting, i.e.\ without resorting to MCMC moves at all.
We re-use the code developed for the original SMC$^2$ paper in order to integrate out the $x_t$ and thus get likelihood estimates, with the same settings for algorithm parameters.
The observed $s_t$ are the $753$ observations from consecutive days of the S\&P $500$ index also used by \citet{chopin2013smc2}.
KGRIS uses a particle system of increasing sizes with each particle going through $100$ iterations.
See Figure~\ref{fig:samples_sv_smc2} in the Appendix for a plot of the pair-wise marginals of this posterior.

We use the same vague priors as \citet{chopin2013smc2},
\begin{gather*}
\mu \sim \DNorm(0,\sigma^2=2), \beta \sim \DNorm(0,\sigma^2=2), \xi \sim \textrm{Exp}(0.2)\\
\omega^2 \sim \textrm{Exp}(0.2), \lambda \sim \textrm{Exp}(1).
\end{gather*}

Figure \ref{fig:eval:sv} shows that the incorporated gradients lead to better performance of KGRIS in estimating the target covariance matrix.
This is in-line with the finding that GRIS improves over pure random walk methods  \citep{schuster2015gradient}.

\section{Discussion}
\label{sec:concl}
In this paper, we developed a framework for kernel sequential Monte Carlo. KSMC adaptively learns the target geometry via kernel emulators and subsequently uses this information for local proposals.
KSMC is especially attractive in the case where likelihoods and gradients are intractable.
We instantiated two algorithms within KSMC:
estimating nonlinear covariance in combination with MCMC rejuvenation and estimating gradients in combination with importance sampling proposals.
Both significantly outperform state-of-the-art gradient-free SMC algorithms in practice.
We conclude with some discussion on computational complexity, more general usage of the learned emulators, 
and on the relative benefits of PMC in the kernel setting.

\paragraph{Computational costs \& increasing dimensions.}
While adaptive schemes for SMC (and MCMC) can increase statistical efficiency of the sampling scheme, they impose additional computational costs.
Somewhat surprisingly, however, these relatively large costs do not severely impact the efficiency per runtime ratio in practice.
The reason is that in the context of intractable likelihoods, the computational cost of fitting a kernel emulator is typically dominated by the larger cost of evaluating model likelihood.
In our real-world experiments on GP classification and a stochastic volatility model in Sections \ref{exp:gp} and \ref{exp:stoch_vol}, a profiler reveals that less than $5\%$ of the overall wall-clock time is spent in computing kernel informed proposals.
This effect increases with dataset size and model complexity, as evaluating likelihood gets more costly.
Clearly however, in the case where we need not resort to pseudo-marginal or SMC${}^2$ type samplers, the application of kernel based estimators might result in slower sampling without much gain in Monte Carlo error.

In growing dimensions, the number of data required to sufficiently estimate nonlinear covariance and gradients quickly becomes infeasible. 
High dimensional sampling problems typically arise in non-parametric models, e.g. Gaussian processes, where each data point comes with additional parameters.
In the intractable likelihood framework that we consider here, however, the marginal posterior over hyper-parameters typically is independent of such latent variables --- and therefore usually of moderate dimension.
Random walk methods, which are the default choice for intractable likelihoods, scale badly in high dimensions themselves \citep{Neal2011}.
Our method is an improvement in the intermediate case: closed form gradients are not available, but the dimensionality of the problem allows to estimate the target geometry just accurately enough to improve mixing.
\citet{strathmann2015gradient} reported their gradient estimator to scale up from dozens to a hundred dimensions on laptop computers, depending on smoothness properties of the target.
It is an active area of research to further scale up these techniques by exploiting structure in the target density.

\paragraph{Emulators as a posterior approximation.}
The kernel approximation of the target density could be considered itself as an output of our algorithms,
representing the posterior directly instead of using the kernel approximation within a sampler.
There are a number of problems with this approach though:
firstly, we note that our emulator models do not need to be perfect to generate useful proposals, therefore allowing us to exploit posterior structure much earlier (even with non-perfect model fit) during sampling, still resulting in a correct SMC sampler.
Also, approximating integrals of test functions with respect to the posterior using the kernel approximation is not possible in closed form, while it is straight forward using a Monte Carlo sum. For example, assume a log density model $f(x)=\sum_i \alpha_i \PDK(x_i, x)$. For the Gaussian kernel $\PDK(x,y)=\exp(-||x-y||^2)$, the density is the exponential of a sum of Gaussian centred at the points $x_i$. Computing an integral as simple as the posterior mean, $\mu = Z^{-1} \int x  \exp(\sum_i \alpha_i \exp(-||x_i - x||^2)) \mathrm{d}x$, already is intractable, even if the evidence $Z$ were known.
Thirdly, it is not possible to sample from the kernel emulator directly using ordinary Monte Carlo. One could imagine running a second MCMC/SMC targeting the emulator model. Not only would this defeat the purpose of the algorithm (this is the problem we are trying to solve in the first place), it also leads to samples that are not guaranteed to consistently estimate posterior expectations unlike kernel SMC or kernel MCMC.

	\paragraph{SMC versus PMC for kernel based proposals.}
	The consensus in the wider SMC community is that using an artificial sequence of proposal distributions for sampling from a static target is preferable to the PMC approach. 
	This is based on the fact that the coverage of the final target is better in these tempering-style algorithms. 
	It however results in a considerable computational investment for those iterations where an intermediate target is considered.

	We also note that on-line updates of the kernel emulator are not possible: the target changes in every iteration. 
	The contrary is true in PMC, where the the actual distribution of interest is targeted in every iteration.
	Here, a popular approximation technique of kernels is a good fit: By expressing the emulator model in terms of finite dimensional random Fourier features, we can perform cheap on-line updates \cite{strathmann2015gradient}. The emulator therefore can accumulate information from all PMC iterations without the computational efforts of re-computing its solution,
providing a relative advantage to SMC in this context. 

\paragraph{\bf Acknowledgments.} I.S.\ was supported by a PSL postdoc grant and DFG through grant CRC 1114 "Scaling Cascades in Complex Systems", Project B03 "Multilevel coarse graining of multiscale problems". H.S.\ was supported by the Gatsby Chaitable foundation. B.P.\ was supported by The Alan Turing Institute under the EPSRC grant EP/N510129/1.

\vspace{-0.1em}
\bibliographystyle{abbrvnat-simple}
{\small
\bibliography{library}
}
\appendix
\newpage

	\section{Implementation details}
	\label{sec:impl}
	In this section, we cover a number of implementation details for using KASMC in practice, such as optimal scaling, adaptive re-sampling and re-weighting between iterations.
	
	\subsection{Scaling parameters}
	\label{sec:scaling}
	Similar to other MH proposals, KAMH has a free scaling parameter denoted $\nu^2$ which we would like to adapt after one SMC iteration.
	To accomplish parameter tuning, we use the standard framework of stochastic approximation for tuning MCMC kernels \cite{Andrieu2008}, i.e.\ tuning acceptance rate $\alpha_t$  towards an asymptotically optimal acceptance rate  $\alpha_\textrm{opt}=0.234$ for random walk proposals \cite{Rosenthal2011}.
	More precisely, after the MCMC rejuvenation step, a Rao-Blackwellised estimate $\hat\alpha_t$ of expected acceptance probability is available by simply averaging the acceptance probabilities for all MH proposals.
	Then, set
	\begin{equation}
	\label{eq:scale_adapt}
	\nu^2_{t+1}=\nu^2_t + \lambda_t(\hat\alpha_t-\alpha_\textrm{opt})
	\end{equation}
	for some non-increasing sequence $\lambda_1,\dots,\lambda_T$.
	This strategy of approximating optimal scaling assumes that consecutive targets are close enough so that the acceptance rate when using $\nu^2_t$ to target $\targd_t$ provides information about the expected acceptance rate when using $\nu^2_t$ with target $\targd_{t+1}$.
	As an alternative to this, one could treat $\nu_t^2$ as an auxiliary random variable and define a distribution over it designed to maximise expected utility, an approach taken in the adaptive SMC sampler \cite{Fearnhead2013}.
	
	\subsection{Construction of a target sequence}
	One possibility for constructing a sequence of distributions is the geometric bridge defined by $$\targd_t \propto \targd_0^{1-\rho_t} \targd^{\rho_t}$$  for some initial distribution $\targd_0$, where $(\rho_t)_{t=1}^T$ is an increasing sequence satisfying $\rho_T = 1$.
	This is the construction used in the experimental section.
	Another construction is to use a mixture $\targd_t \propto ({1-\rho_t})\targd_0+ {\rho_t}\targd$.
	When $\targd$ is a Bayesian posterior, one can also add more data with increasing $t$, e.g. by defining the intermediate distributions as $\targd_t(\smp) = \targd(\smp|d_1,\dots,d_{\floor{\rho_t D}})$ where $d_j$ is a datapoint and $D$ is the number of data points.
	This results in an online inference algorithm called Iterated Batch Important Sampling (IBIS) \cite{Chopin2002}.
	In IBIS especially, we can apply non-diminishing adaptation, unlike in adaptive MCMC.
	
	When using a distribution sequence that computes the posterior density $\targd$ using the full dataset (such as the geometric bridge or the mixture sequence), one can reuse the intermediate samples when targeting $\targd_t$ for posterior estimation.
	As the value of $\targd$ is computed for the geometric bridge and the mixture sequence, we re-use the weight $\targd(\smp)/\targd_{t-1}(\smp)$ for posterior estimation while employing $\targd_t(\smp)/\targd_{t-1}(\smp)$ to inform proposal distributions at iteration $t$.
	This way, the evaluation of $\targd$ (which is typically costly) is put to good use for improving the posterior estimate. 
	
	As a simple alternative, leading to the algorithm known as Population Monte Carlo (PMC) \cite{Cappe2004}, 
	we can simply target the final distribution $\pi$ at each iteration, i.e.~with all $\pi_t = \pi$.
	The original work on PMC exhibited striking resemblance of  commonly used MCMC methods such as Random Walk metropolis, often finding that the same proposal kernel with PMC produces better estimates than with MCMC \cite{Cappe2004}.

	\subsection{Re-weighting and adaptive re-sampling}
	The fact that the weighted approximation to the final target is returned in our algorithm stems from the fact that this approximation has lower variance than the re-sampled particle system \cite{Doucet2009}.
	This is why in practice re-sampling might not be performed at every iteration.
	Rather, re-sampling only when  Effective Sample Size (ESS) for the current target falls below a certain threshold will decrease Monte Carlo variance.
	For details we refer to reviews on SMC \cite{Doucet2009,Doucet2001}.
	Furthermore, care should be taken with respect to implementation of re-weighting: caching values between iterations saves much computation time.

	\subsection{Intractable Likelihoods and Evidence Estimation}
	\label{sec:evidence_estimation_noisy_likelihoods}

	In the the case where likelihoods are intractable, SMC is still a valid algorithm when likelihood values can be estimated unbiasedly. 
	This can be done using e.g.\ importance sampling or SMC \cite{Tran2013,chopin2013smc2}.
	A simple way to think about such nested estimation schemes is in terms of an extended sampling space that spans the actual parameters of interest as well as any nuisance variables.
	Intractable likelihoods usually result in unavailability of gradients.
	Consequently, efficient gradient-based sampling schemes based such as GRIS or HMC are unavailable.
	Current practice there is based on moving particles using random walk schemes solely.
	
	An important issue in Bayesian model selection and averaging is that of estimating the normalizing constant, or \emph{evidence}.
	The evidence is the marginal probability of the data under a model and can easily be estimated in SMC instantiations \cite{Doucet2009,Fearnhead2013} -- as compared to MCMC.
	This enables routine computation of Bayes factors and posterior model probabilities while also sampling from a posterior over parameters of each model.
	Under the assumption that the normalizing constant $\evid_0$ of $\targd_0$ (the distribution that is used for initially setting up the particle system) is known, one can estimate the ratio of normalizing constants of any two consecutive targets by
	\begin{equation}
	\label{eq:evid_ratio_smc}
	\frac{\evid_t}{\evid_{t-1}} \approx \frac{1}{\psize} \sum_{j=1}^\psize \iw_{t,j}
	\end{equation}
	for $\iw_{t,j} = {\targd_t(\smp_{t-1,j})}/{\targd_{t-1}(\smp_{t-1,j})}$ and thus an estimate for $\evid = \evid_T$ can be found recursively by
	\begin{equation}
	\label{eq:evid_approx_smc}
	\evid = \evid_T \approx \evid_0\prod_{t=1}^T \frac{1}{\psize}\sum_j \iw_{t,j}
	\end{equation}
	starting with known value $\evid_0$.
	When the likelihood is intractable and importance weights are noisy, evidence estimation is still valid  \citep[Lemma 3]{Tran2013}.

	\begin{figure}[b]
		\centering
		\includegraphics[width=\textwidth]{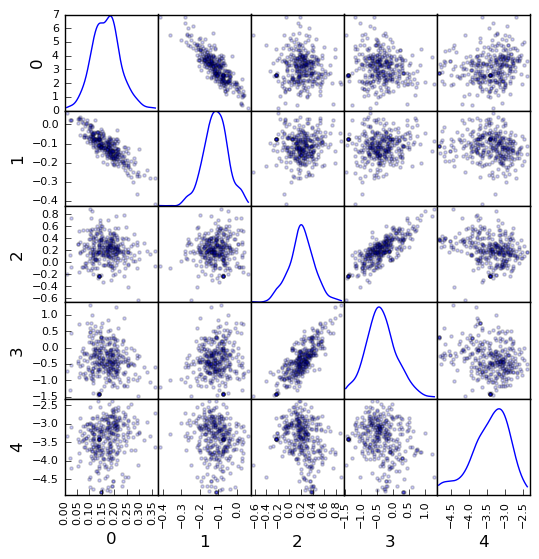}
		\caption{Samples from the doubly stochastic volatility model used in Section \ref{exp:stoch_vol}.}
		\label{fig:samples_sv_smc2}
	\end{figure}

\end{document}